# Analysis and design of a Germanium multi-quantum well metal strip nanocavity plasmon laser

HAMED GHODSI,[1] HASSAN KAATUZIAN,[1,*] AND ELAHE RASTEGAR PASHAKI[1]

[1]*Photonics Research Laboratory (PRL), Electrical Engineering Department, Amirkabir University of Technology*
*\*hsnkato@aut.ac.ir*

**Abstract:** In order to achieve electrically pumped plasmon nano lasers, several structures, materials and methods, have been proposed recently. However, there is still a long way to find out a reliable appropriate on-chip plasmon source for commercial plasmonic integrated circuits. In this paper, a new waveguide integrated nanocavity plasmon laser is proposed for 1550 nm free-space wavelength. Due to its significant field confinement resulted by the metal strip structure and strong interaction of plasmonic modes with the germanium quantum wells and as a result a considerable Purcell factor about 291, this structure has a remarkable output performance. Using semi-classical rate equations in combination with finite difference time domain (FDTD) cavity mode analysis, the output performance measures are estimated and confirmed with respect to various physical models and simulation tools. Simulation results for this tiny structure (0.073 µm$^2$ area) show a 2.8µW output power with 10µA injection current and about 4.16mW output power with the threshold pump current of 27mA while maintaining its performance in a wide modulation bandwidth of 178GHz.

## 1. Introduction

A room temperature electrically pumped integrated plasmon source that generates output power in the mW range is the key component for the realization of future integrated plasmonic chips. Providing a practical solution can revolutionize photonics and optoelectronics into all plasmonic integrated circuits, which have significant developments for various plasmonic counterparts of photonics and optoelectronic devices in recent years. [1-5] To do so, several numbers of researches on design and fabrication of plasmon sources i.e. plasmon nanolasers or SPASERs were done. These plasmon sources can be categorized in metallic nanoshells [6], nanocavities [7], nanowires [8] and waveguide-based nanolasers [9]. As mentioned, there are many laboratory fabrications and theoretically proposed devices, which in many cases have optical pumping or cryogenic operating conditions or complicated fabrication process [6,8]. Therefore, these devices are not ready for this purpose yet and there is still demand for practical plasmonic nanolasers.

In this paper, an electrically pumped Germanium/Silicon-Germanium (Ge/Si$_{0.11}$Ge) multiple quantum well plasmonic nanolaser is introduced, analyzed and simulated. The proposed nanolaser has a thin Gold metal strip structure, sandwiched between Ge quantum wells in order to maximize both field confinement and exciton-plasmon interaction possibility, which means higher Purcell and better gain medium/resonator mode overlap factors. This design is covered by two Aluminum electrical contacts for applying the pump current. In addition, it can be coupled into both MIM and IMI silicon-based plasmonic waveguides with considerable coupling efficiency or used in far-field configuration in which plasmon modes will be converted into photons through the cavity interface. This device benefits from a Metal-Semiconductor-Metal-Semiconductor-Metal (MSMSM) structure, which can perform well in 1550nm regime by means of incorporating highly doped strained Ge quantum wells as the direct bandgap gain medium. [10]

In the next section, physical structure and fabrication considerations are explained. Then in section 3, governing physical principles are explained and numerically calculated. Following this, in section 4, the results of output characteristics analysis can be seen for different output couplings and this paper is concluded in the fifth section.

## 2. Physical structure and Fabrication

As can be seen from Fig.1 the 3D structure of our proposed nanolaser consists of four different materials including Al top and bottom contacts, Germanium, and $Si_{0.11}Ge$ Buffer layers, strained Ge quantum wells and thin Gold strip in the middle forming the MSMSM nanocavity structure. Furthermore, different coupling scenarios are proposed and can be witnessed from Fig.2.A, Fig.2.B, and Fig.2.C respectively.

The proposed nanolaser would be fabricated on a Silicon substrate as follows. It should be mentioned that we are not going to describe the fabrication process in detail; instead, in each fabrication step, an empirically fabricated similar structure will be referenced. A more detailed lateral structure and the design specifications are provided in Fig.3 and Table.1 respectively.

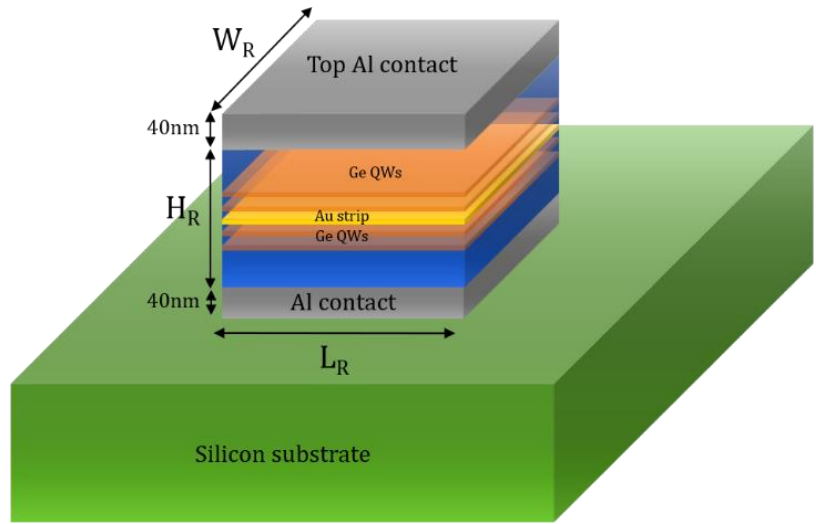

Fig.1 3D schematics of proposed nanolaser

The fabrication process can be started on a silicon substrate by deposition of the Aluminum bottom contact as stated in [11], followed by deposition of the Germanium and $Si_{0.1}Ge_{0.9}$ bottom buffer layers and then two Germanium quantum wells. [12,13] This process will be continued by deposition of the Gold metal strip, which can be taken from [13] and again a similar process to the bottom Ge/SiGe layers [12,13] and finally deposition of top Aluminum contact. [11] The fabricated structure can then be connected to a metal/insulator/metal (MIM) or insulator/metal/insulator (IMI) waveguide or even be used in far field lasing mode. (Fig.2)

Our proposed nanolaser performs magnificently due to its strong mode confinement and tiny modal volume. In addition, significant overlap of plasmonic mode with the quantum wells results in higher output power and finally, two-sided mode profile doubles up the output. This great performance is a tradeoff for the complicated fabrication process and increased plasmon loss.

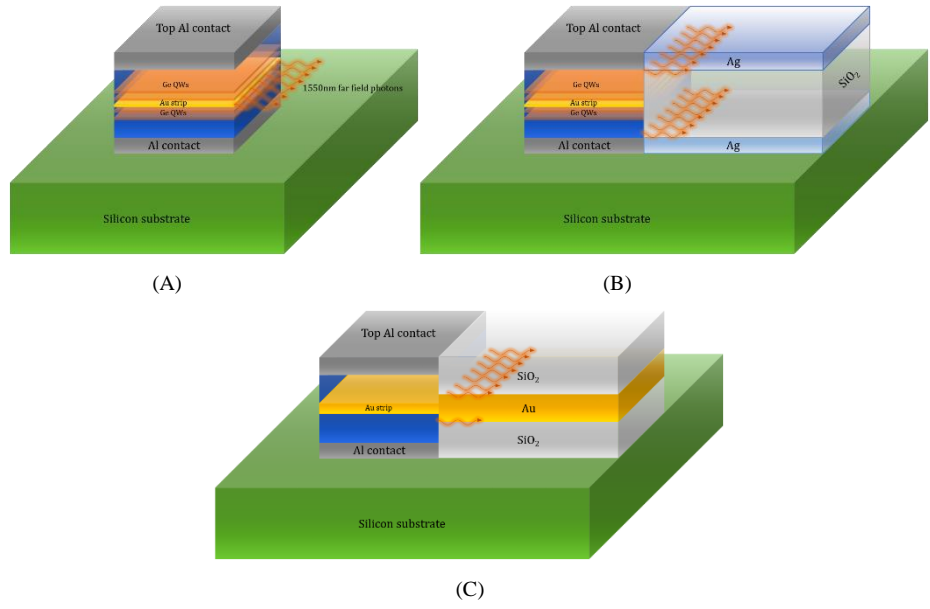

(A)

(B)

(C)

Fig. 2 Proposed output coupling configurations. A. Far-field configuration. B. Coupling to MIM waveguide. C. Coupling to IMI waveguide.

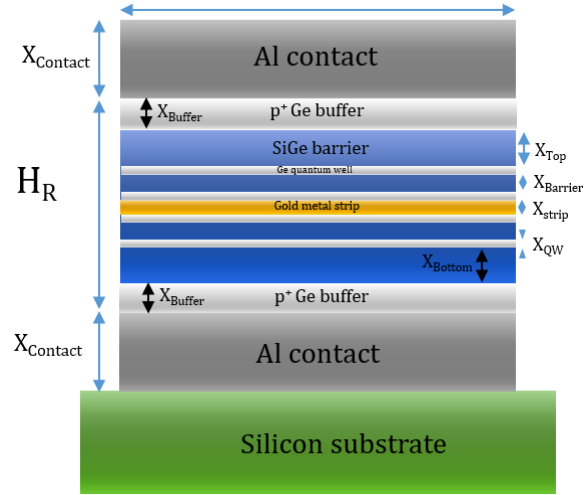

Fig. 3 detailed lateral structure of the Cavity

Table 1. Design parameters of the proposed nanolaser

| Symbol | Description | Value | Unit |
| --- | --- | --- | --- |
| $W_R$ | Resonator size | 270 | nm |
| $H_R$ | Resonator height | 130 | nm |
| $X_{Strip}$ | Bottom metal thickness | 10 | nm |
| $X_{Contact}$ | Top metal thickness | 40 | nm |
| $X_{Bottom}$ | Bottom buffer thickness | 15 | nm |
| $X_{Top}$ | Bottom buffer thickness | 15 | nm |
| $N_{QW}$ | Number of QWs | 4 | - |
| $X_{QW}$ | QW thickness | 7 | nm |
| $X_{Barrier}$ | Barrier wall thickness | 10 | nm |
| $X_{Buffer}$ | Thickness of p doped Ge buffer | 16 | nm |
| $x$ | Ge Alloy percent | 89 | % |
| $N_D$ | Doping concentration of the QWs and Barriers | $7.6 \times 10^{19}$ | cm$^{-3}$ |
| $N_A$ | Doping concentration of the Ge Buffer | $1 \times 10^{19}$ | cm$^{-3}$ |

## 3. Governing Theories

Following the approach same as a traditional laser, in a plasmonic one, we also should have gain medium and a resonator. First, we will discuss resonator and mirrors and then we will explain the gain medium equations. Finally, we will be able to write the rate equations for simulation of the output power, pumping threshold, plasmonic gain and modulation bandwidth in the next section.

### 3.1. Cavity characterization and output coupling

The proposed metal strip plasmonic nanocavity can be characterized by its resonance wavelength, quality factor, equivalent modal volume, Purcell factor, and confinement and coupling factors.

The resonance wavelength can be found by calculating propagation modes decay rate versus frequency in the resonator which can be witnessed in Fig.4 for different cavity sizes while other parameters were taken from Table.1. It is worth mentioning that the complex dielectric constants are taken from CRC model of [14] for Gold, Aluminum, and Germanium and for SiGe a linear interpolation is done between dispersion models of [14]. From this analysis for cavity size of 270nm, the resonance frequency is calculated to be 194.31THz which is equivalent to 1542.86 nm free-space wavelength.

Quality factor can be expressed by (1) for the plasmonic nanowire resonator:

$$Q = 2\pi \frac{\text{Energy stored in cavity}}{\text{Energy lost per cycle to walls}} \quad (1)$$

Using design values of Table.1 and FDTD method, cavity and mirror loss values have been calculated while the first propagation mode of the cavity was applied as the source. (See Figure.5) The "Q" factor is then calculated and its value for 1550nm wavelength and the fundamental mode is about 26.37.

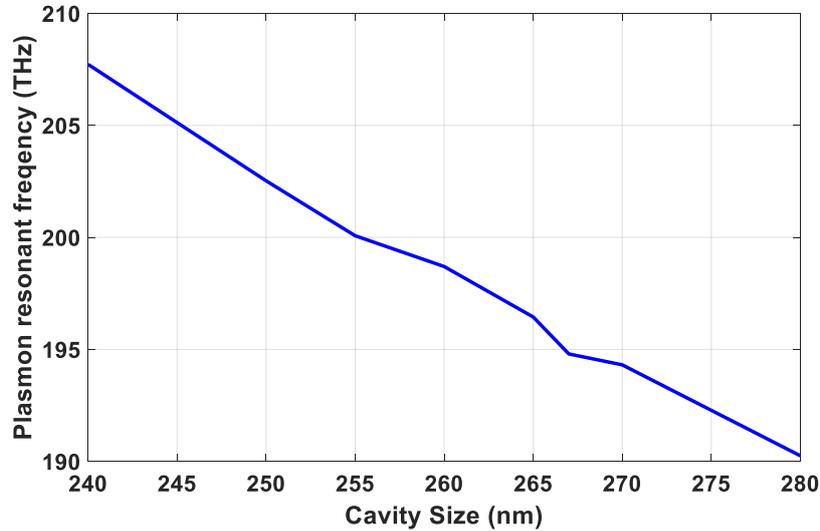

Fig.4 Resonance frequency versus cavity width while other parameters are set from Table.1

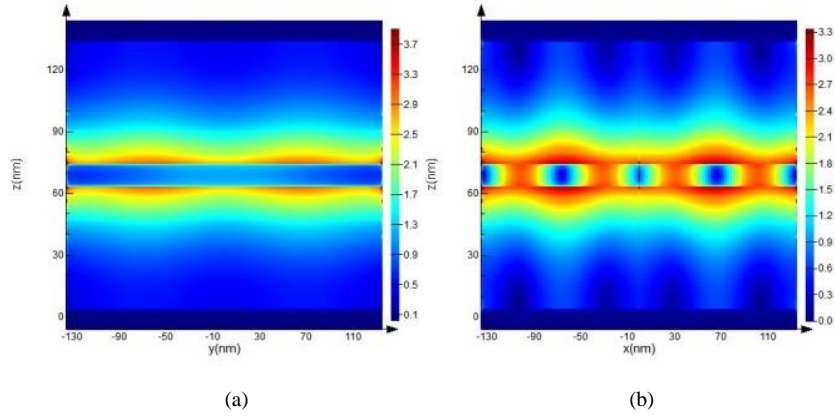

(a)          (b)

Fig.5 Transverse (a) and longitudinal (b) cross sections of the cavity fundamental mode electric field distributions respectively.

In plasmonic metal nanocavities, considering their large amount of loss, the quality factor is far less than their insulator optical counterparts. Therefore, the quality factor of our proposed nanocavity despite having considerable mode confinement and tiny modal volume is very low. In order to increase the quality factor, cavity loss that originates from both scattering and mirror losses should be compensated. Propagation mode loss of the plasmons in the metal strip even by using gold is still high and about $1.07 \times 10^5$ dB/cm in room temperature (300K). This can be significantly lowered by decreasing the working temperature that can lower the loss by a factor of 0.01 for each Kelvin. [15] For instance, in 100K, scattering loss will be halved. In addition, for decreasing the mirror loss a plasmonic Insulator/Metal/Insulator (IMI) Bragg reflectors as shown in Fig.6 can be utilized. The Bragg condition for calculation of the grating period can be written as (2). [16]

$$\frac{2\pi}{\lambda_0}\left(\frac{L_{Period}}{2}\right)\left[n_{eff1} + n_{eff2}\right] = (2m+1)\pi \qquad (2)$$

Where "$\lambda_0$" is free-space wavelength, "$L_{Period}$" is Bragg period, "$n_{eff1}$" and "$n_{eff2}$" are the effective refractive indexes for the thin and thick parts of the waveguide respectively and m is an integer.

For the structure shown in Fig.6, the Bragg period is calculated to be 90nm. According to FDTD simulations for two periods on each side of the resonator, the quality factor is increased approximately by a factor of 2. However, it is a tradeoff between decreasing the mirror loss which results in increased "Q" factor and thus output power and increasing device dimension and lowering output coupling ratio.

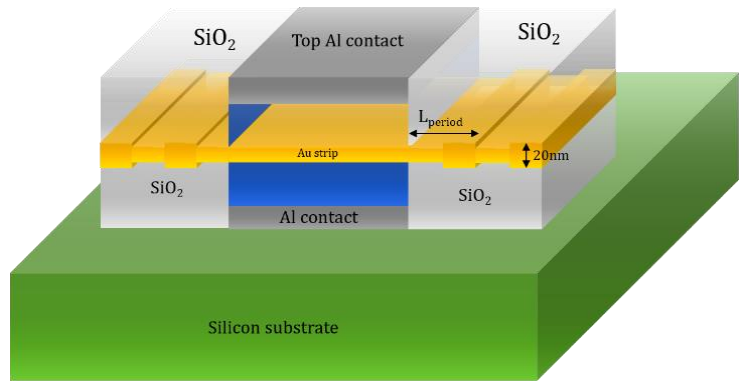

Fig.6 IMI Bragg reflectors for decreasing mirror loss

Effective mode volume "$V_{eff}$" has great importance in nanolaser operation, which can be calculated by (3). [14]

$$V_{eff} = \frac{\int_V \varepsilon(r)|E(r)|^2 d^3r}{\max\left[\varepsilon(r)|E(r)|^2\right]} \quad (3)$$

Where "ε" is dielectric constant, "E" is the electric field and "V" is the resonator volume. From the FDTD analysis using lumerical package for the optimal values of Table.1 equivalent mode volume for 193.54 THz, which is equal to 1550nm free space wavelength is about 3.2797×10$^{-16}$cm$^3$ on each side of the metal strip which is much less than a metal/semiconductor nanocavity of the same size. [7] This significantly high mode confinement and two-sided power propagation can guarantee a considerable output performance.

The Purcell factor [17] "$F_p$", is a key parameter in cavity quantum electrodynamics (CQED) that defines the coupling ratio between a dipolar emitter (QWs in our case) and a cavity mode. Purcell factor as can be expressed in (4) specifies the possible strategies to enhance and control light-matter interaction. [18] Efficient light-matter interaction is achieved by means of either high quality factor "Q" or low modal volume "V", which is the basis of plasmonic cavity quantum electrodynamics (PCQED). [19]

$$F_p = \frac{3}{4\pi^2}\left(\frac{\lambda}{n}\right)^3\left(\frac{Q}{V_{eff}}\right) \quad (4)$$

Where "λ" is free-space wavelength, "n" is the refractive index of gain medium and Q is the quality factor for the plasmonic resonator. There is also an alternative way of finding the Purcell factor. To do so, we have to use two dipole sources on each side of the gold strip for applying the same electric and magnetic fields as the propagation mode. Then the Purcell factor can be calculated by averaging the ratio of the power emitted by a dipole source in the environment to the power emitted by the dipole in a homogeneous environment (bulk material) for the two dipole sources. However, the results were pretty much the same for the two methods in our FDTD simulations and we have used this fact for double checking. Since the emission rate is proportional to the local density of optical states (LDOS), and the LDOS is proportional to the power emitted by the source. [14] Among all of the propagation modes, fundamental mode, as can be witnessed in Fig.5, has a considerable overlap with the quantum wells and the least loss value according to Fig.7. Thus, the fundamental mode is more likely to be excited. The Purcell factor for 193.54THz and for the fundamental mode is equivalent to 291.

β which is known as the coupling factor is defined by the ratio of the spontaneous emission rate into the lasing mode (Fundamental mode) and the spontaneous emission rate into all other modes and can be expressed by (5). [7]

$$\beta = \frac{F_{cav}^{(1)}}{\sum_k F_{cav}^{(k)}} \quad (5)$$

Where $F_{cav}^{(k)}$ is the Purcell factor of the k'th mode. k = 1 corresponds to the lasing mode and the summation is on all cavity modes. For calculating β factor, a method based on several randomly positioned dipole sources is used where the lasing mode is determined by the dipole source with maximum Purcell factor. By means of (5) and calculation of Purcell factors for all of these dipole sources (As shown in Fig.8), the β factor can be determined and it is equal to 0.0685 for the proposed square cavity structure.

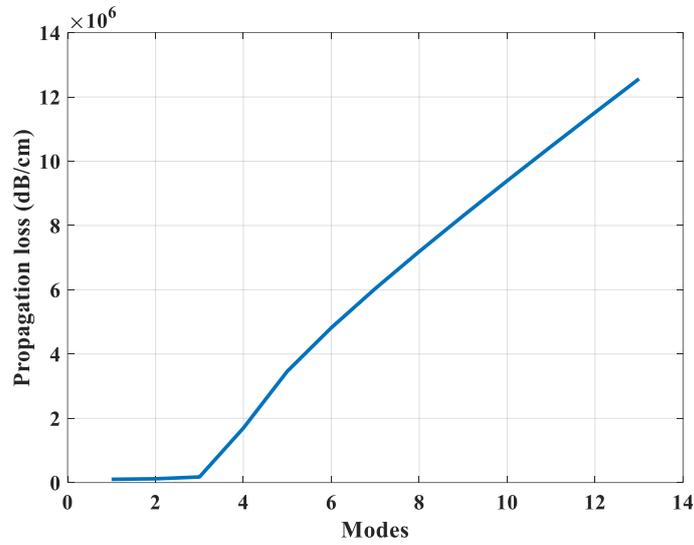

Fig.7 Propagation loss in dB/cm for different modes. Mode "1" is the fundamental mode

It should be noticed that for a mode to be considered as the lasing mode, there are two main measures. Firstly, the mode overlap with the gain medium which is also known as "Γ" factor, should be maximum. "Γ" has been calculated by the dividing the volume in quantum wells in which the electrical field is more than the half of the maximum electrical field onto the volume of the gain medium, i.e. The four quantum wells. The fundamental mode as shown in Fig.5 has the highest overlap factor which is calculated to be about 44.4% while for the other modes as can be seen from the inlet of Fig.8 "Γ" is significantly lower than the fundamental mode. Secondly, the longitudinal mode profile should provide power on the ends to be transmitted outside and locally confined modes even with higher Purcell factors cannot be considered as the lasing mode.

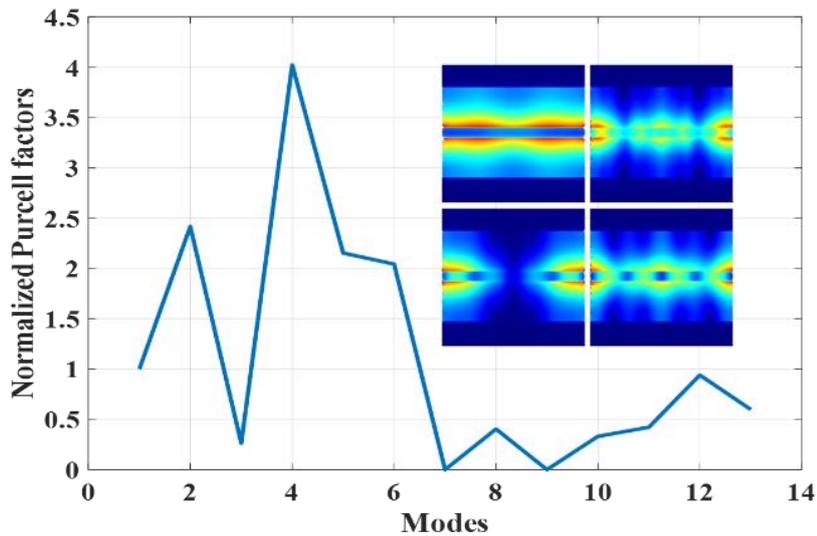

Fig.8 Normalized Purcell factor for some of the resonator modes. The values are Normalized to the Purcell factor of the fundamental mode. (Inlet: Transverse cross section for 4 of the calculated modes)

Coupling factor for the far field scenario of Fig.2.A can be easily calculated by the transmission ratio of the mirrors from Fresnel's equation [20] and in our case, it is about

61.45%. For MIM scenario of Fig.2.B as can be seen in Fig.9 and FDTD analysis, the coupling efficiency is about ~30%. Eventually for IMI case as can be seen from Fig.10 for three different metal thicknesses of the IMI waveguide, the coupling factor to the host waveguide is provided where the maximum coupling factor is about 71.44% for the metal thickness of 10nm.

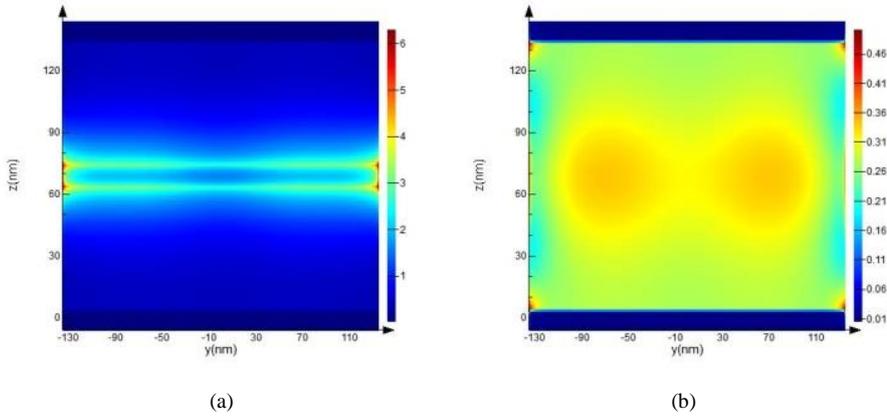

(a) (b)

Fig.9 MIM coupling transverse field profile cross-sections. a. At the cavity edge b. 35nm away from the cavity edge)

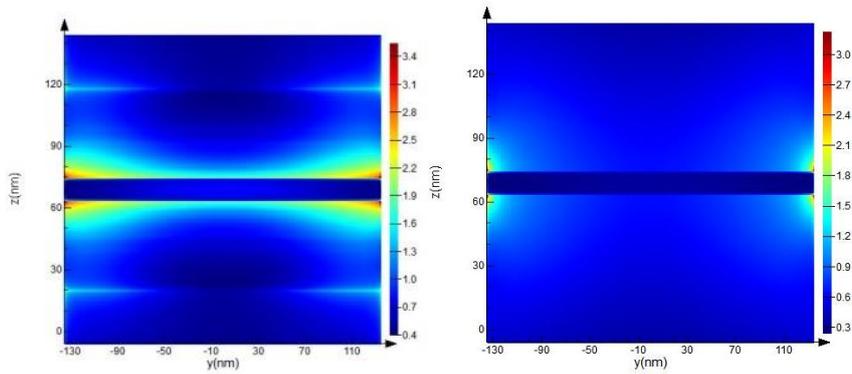

a. 10nm, Coupling ratio = 71.44%

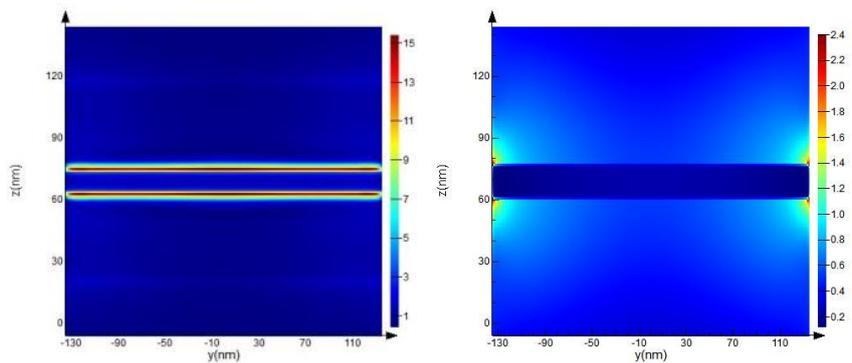

b. 15nm, Coupling ratio = 67.51%

Fig.10 IMI coupling ratio for different metal thicknesses at the edge and 35nm away (a. 10nm b.15nm c. 20nm)

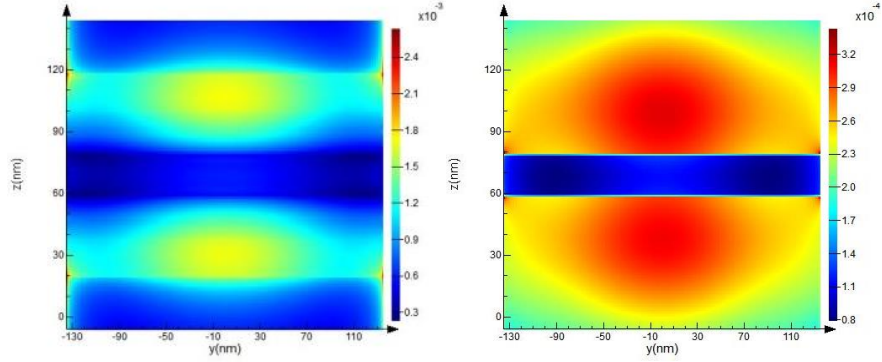

c. 20nm, Coupling ratio = 34.29%

Fig.10 IMI coupling ratio for different metal thicknesses at the edge and 35nm away (a. 10nm b.15nm c. 20nm)

### 3.2 Rate equations and energy transfer

In the proposed structure as can be seen in Fig.11, the energy of generated excitons in quantum wells due to electrical current will be transferred to Surface Plasmon Polariton (SPP) modes in both sides of the gold strip. In order to analyze the performance of a plasmonic nanolaser, we need a model for its rate equations. For this purpose, we will start with a set of semi-classical rate equations similar to the microcavity semiconductor lasers which can be witnessed in (6). [21] In these set of equations, the first one is expressing the rate of carrier changes and the second one is expressing the temporal changes in the plasmon generation which is affected by spontaneous plasmons coupled in the lasing mode (the first term), stimulated emission (the second term) and plasmon loss (the last term). [21]

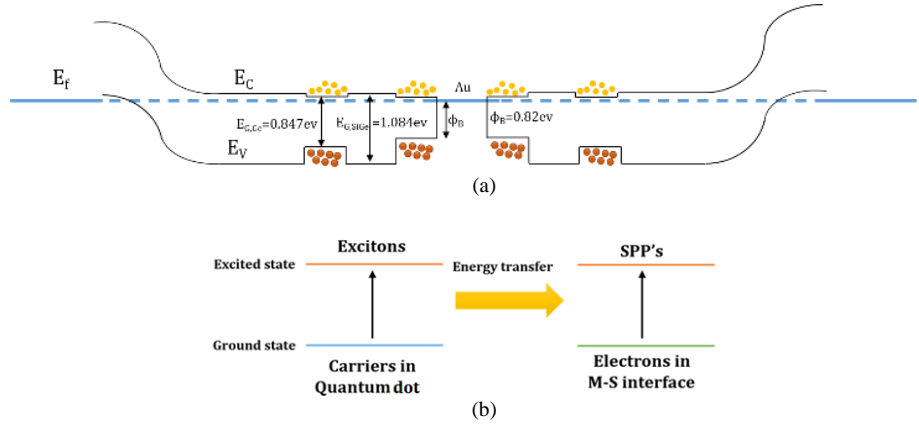

Fig.11 Energy transfer diagram. a. Energy band diagram of nanolaser in equilibrium, b. Energy transfer concept

$$\frac{dn}{dt} = P - An - \beta \Gamma As(n-n_0) - \frac{nv_s S_a}{V_a}$$

$$\frac{ds}{dt} = \beta An + \beta \Gamma As(n-n_0) - \gamma s$$

(6)

In these equations "n" is the excited state population of the carriers, "s" is the number of plasmons in the lasing mode and "P" is the carrier generation rate in "$s^{-1}$". Total carrier generation rate (P) is determined by several parameters like, pump current (rate of injected carriers by electrical pumping), thermionic emission over and tunneling rates through Schottky barrier (in this case is neglected due to the very high doping level), contact to QWs transit time (drift/diffusion theory [22]), transition probability from each well, carrier recombination lifetime in quantum wells (indicates the average time before an exciton transfers its energy to the SPP lasing mode or lose energy due to other processes) which can be expressed by radiative recombination rate and non-radiative recombination rates [10] (Auger and SRH processes) and tunneling and thermionic emission probability between two neighbor quantum wells. All of the mentioned phenomena should be considered to achieve a precise model for finding pump rate (P) as a function of pump current (a carrier dynamics model). In this paper, we have proposed a simplified model, which is based only on direct and indirect recombination coefficients. In this model, internal quantum efficiency (η) can be calculated from (7).

$$R_{Auger} = C_{nnp}n + C_{ppn}p$$
$$R_{nonradiative} = R_{Auger} + R_{SRH}$$
$$R_{total} = R_{radiative} + R_{nonradiative} \quad (7)$$
$$\eta = \left(\frac{R_{radiative}}{R_{total}}\right)$$

Where "$R_{radiative}$" is radiative recombination coefficient, which is equal to "$1.3 \times 10^{-10}$ cm$^3$/s" for Ge quantum wells and "$R_{nonradiative}$" is non-radiative recombination coefficient respectively. [10] For "$C_{nnp}$" and "$C_{ppn}$" which are equal to "$3 \times 10^{-32}$ cm$^6$/s" and "$7 \times 10^{-32}$ cm$^6$/s" respectively and SRH recombination rate which is equal to "$5.1 \times 10^{-15}$ cm$^3$/s", non-radiative recombination rate in our device is equivalent to "$2.285 \times 10^{-12}$ cm$^3$/s". [10] As a result internal quantum efficiency is "98.24%" which means radiative recombination is totally dominant.

Due to the symmetric structure, calculations were done for each side of the metal waveguide individually and then the output power has been calculated by superposition of top and bottom plasmon generations. Therefore, the carrier generation rates for the top "$P_1$" and bottom sides "$P_2$" can be expressed as a function of injection current as can be seen in (8).

$$P_1 \simeq P_2 = \frac{N_{QW}V_{QW}}{2}R_{radiative}(np - n_i^2) \rightarrow P_1 \simeq P_2 \simeq \frac{N_{QW}V_{QW}}{2}R_{radiative}N_D\Delta p$$
$$\Delta p = \left(\frac{I_{pump}}{qN_{QW}V_{QW}}\right)\tau_c \rightarrow P_1 \simeq P_2 \simeq \frac{I_{pump}\tau_c R_{radiative}N_D}{2q} \quad (8)$$

Where, "$I_{pump}$" is injection current in "A", "n" is majority electron concentration in "cm$^{-3}$", "p" is minority hole concentration in "cm$^{-3}$", "$n_i$" is intrinsic carrier concentration, "q" is electron charge in "C" and "$\Delta p$" is the steady-state population of excess holes in the gain medium. In addition, "$\tau_c$" is the effective recombination lifetime, which can be estimated by (9) in which phonon-assisted carrier capture and escape times [23] are neglected. In our case P in "s$^{-1}$" is related to the injected current in "A" by a coefficient of $2.75 \times 10^{16}$ A$^{-1}$·s$^{-1}$.

$$\tau_c^{-1} = \tau_{QW_1}^{-1} + \tau_{QW_2}^{-1} + \tau_{QW_3}^{-1} + \tau_{QW_3}^{-1}$$
$$\tau_{QW_1} = \tau_{Al-QW_1} + \tau_{rec}$$
$$\tau_{QW_2} = \tau_{Al-QW_1} + \tau_{w-w} + \tau_{rec} \quad (9)$$
$$\tau_{QW_3} = \tau_{Al-QW_1} + \tau_{w-w} + \tau_{Au} + \tau_{rec}$$
$$\tau_{QW_4} = \tau_{Al-QW_1} + 2\tau_{w-w} + \tau_{Au} + \tau_{rec}$$

Where "$\tau_{Al-QW1}$" is the transit time of the carriers from the contact to the first quantum well which is determined by saturation velocity of the electrons due to the compact size of the nanolaser and for 31nm buffer layer (16nm p-doped Ge and 15nm n-doped SiGe) of our structure "$\tau_{Al-QW1}$" is about 0.33ps. "$\tau_{rec}$" is the recombination lifetime in one of the Germanium quantum wells which can be calculated by the total recombination coefficient "$R_{total}$" and assuming one excess carrier in the quantum well region and it is about 0.73ps. Inter-quantum well transition mechanism as expressed by "$\tau_{w-w}$" is determined by "$(\tau_{tunnel}^{-1} + \tau_{therm}^{-1})^{-1}$" where "$\tau_{tunnel}$" and "$\tau_{therm}$" are related transit times of tunneling through and thermionic emission over the barrier mechanisms respectively. [24] In our device according to (10) the inter-quantum well transit time is dominated by the tunneling process and it is about 13ps. [24]

$$\tau_{tunnel} = \frac{m_b^* X_{qw}^2 \left(1 + GX_{qw}\right) e^{Gl_b}}{2h}$$

$$\tau_{therm} = X_{QW} \sqrt{2\pi m_{qw}^*} \times \frac{e^{\frac{E_{G,SiGe} - E_{G,Ge}}{kT}}}{\sqrt{kT}}$$

(10)

Where "$m_b^*$" and "$m_{qw}^*$" are effective mass of the electrons in the barrier region and quantum wells respectively, "$W_{qw}$" is the quantum well thickness, "h" is the Plank's constant, "$l_b$" is the barrier thickness "k" is Boltzmann constant, T is temperature in Kelvin, "$E_{G,SiGe}$" and "$E_{G,Ge}$" are Barrier and quantum well bandgaps and eventually "G" is the energy-independent wave number which is 6451.61cm$^{-1}$ for 1550nm. In addition, "$\tau_{Au}$" is drift time of the electrons in the 10nm Gold strip which is in femtosecond scale and neglected in comparison to other mentioned time constants. Accordingly, "$\tau_c$" can be calculated and it is about 891.6fs.

"A" is the spontaneous emission rate, which can be modified by the Purcell effect via "$A = F_p A_0$", where "$A_0$" is the natural spontaneous emission rate of the material equals to $1/\tau_{sp0}$ and $\tau_{sp0}$ is the spontaneous emission lifetime of the gain medium which is equal to "$\tau_c/\eta$" that results in 907.6fs spontaneous emission lifetime.

"Γ" which equals to the ratio of carriers generated in the spatial distribution of plasmonic modes to the whole number of generated carriers, is also called mode overlap with the gain medium coefficient and as calculated before, it is 0.444. "$n_0$" is the excited state population of carriers in transparency and it is about 3.5×10$^{18}$cm$^{-3}$, [10] "$v_s$" is surface recombination velocity at the sidewalls of the resonator which equals to 2160 cm/s [25]. "$S_a$" and "$V_a$" are the area of sidewalls of the nanolaser and volume of gain medium, which are equivalent to 1.188×10$^{-10}$ cm$^2$ and 1.0206×10$^{-15}$ cm$^3$ respectively. Eventually, "γ" is the total loss rate of plasmons in the cavity (loss coefficient per unit length × modal speed), which is calculated by "$\gamma_m + \gamma_g$" where "$\gamma_m$" and "$\gamma_g$" are resonator mirror loss and loss due to the gain medium per unit length respectively. Loss due to gain medium will be calculated by integrating the imaginary part of metal permittivity in the desired frequency along the path of SPPs and loss due to mirrors will be calculated by Fresnel's law [19]. Accordingly, "γ" is 4.151 fs$^{-1}$.

In order to compare the output performance of plasmon lasers, various figures of merit can be considered. Among these, we have used output power, threshold pump current, plasmonic gain, and operational bandwidth.

Output power as can be witnessed in (11) is a function of the number of generated plasmons in a plasmon lifetime in the cavity and can be derived from the rate equations of (6). [7]

$$P_{out} = \eta_c \times \frac{\gamma_m}{\gamma_m + \gamma_g} \times \frac{S}{\tau_p} \times \frac{hc}{\lambda}$$

(11)

Where "$\eta_c$" is the coupling efficiency "S" is the number of generated plasmons in a plasmon lifetime, "$\tau_p$" is plasmon lifetime in the cavity and equals to "$Q \times (2\pi f_{res})^{-1}$" in which "Q" is the quality factor and "$f_{res}$" is the resonance frequency of the cavity, "h" is Planck's constant, "c" is light speed and "$\lambda$" is the output wavelength.

Plasmonic gain as expressed by (12) can be defined by the increase in the number of plasmons per unit length of the cavity. [20]

$$G_{spp} = \frac{\beta \Gamma A \int_{\tau_p} s(n-n_0)dt}{v_g \int_{\tau_p} s} \quad (12)$$

Where "$G_{spp}$" is the plasmonic gain in "$cm^{-1}$" and "$v_g$" is the group velocity of the SPPs in "cm/s" which is expressed in (13) and in our device is about "$4.44 \times 10^9 cm/s$".

$$v_g = \frac{d\omega}{d\beta_{spp}}, \beta_{spp} = \frac{\omega}{c}\sqrt{\frac{\varepsilon_{Ge}\varepsilon_{Au}}{\varepsilon_{Ge}+\varepsilon_{Au}}} \quad (13)$$

Where "$\omega$" is the SPP frequency in "rad/s", "c" is the speed of light in vacuum in "m/s" and "$\varepsilon_{Ge}$" and "$\varepsilon_{Au}$" are the dielectric constants of Germanium and Gold respectively.

Threshold pump current as can be calculated by (14) is the point in which plasmonic gain of the nanolaser is equal to the total loss per unit volume of the cavity consisting both mirror loss "$\gamma_m$" and loss due to the gain medium "$\gamma_g$". Accordingly, the threshold pump rate is $7.69 \times 10^{14} s^{-1}$ and threshold current is about 29mA.

$$P_{th} = P\big|_{G_{spp}=\gamma_m+\gamma_g} \rightarrow I_{th} = \frac{P_{th}}{2.75 \times 10^{16}} \quad (14)$$

The bandwidth of the proposed nanolaser is characterized by three main time constants as can be seen in (15).

$$BW = \frac{1}{2\pi(\tau_c + \tau_{cap} + \tau_{spp})} \quad (15)$$

The first time constant is the effective recombination lifetime and as mentioned before it can be estimated by (9) to be 891.6fs. Parasitic RC time constant of the nanolaser, "$\tau_{cap}$" can be calculated for "$R_s$" a standard "50Ω" resistance of the modulation source by (16).

$$\tau_{cap} = \frac{\left(R_s \varepsilon_c \frac{W_R^2}{(H_R - X_{strip})}\right)}{2} \quad (16)$$

Where "$\varepsilon_c$" is the average dielectric constant of both Germanium QWs and SiGe barriers and "$\tau_{cap}$" is equal to "0.37fs" for our device. The third parameter, "$\tau_{spp}$" which contributes for SPP dynamics that can be calculated from 3dB bandwidth of spectral response transfer function of (17) in the threshold pumping condition is equal to "3.33fs". [7]

$$\tau_{plasmon} = \frac{1}{\omega_{3dB}}$$

$$\omega_{3dB} : \omega \ @ \ H(\omega) = \frac{1}{2} H(0) \tag{17}$$

$$H(\omega) = \frac{\beta A (1 + S_0)}{\sqrt{\left(\omega^2 - \omega_r^2\right)^2 + \omega^2 \omega_p^2}}$$

Where "$\omega_r$" and "$\omega_p$" are derived from (18) and (19) respectively and "$S_0$" is the steady-state plasmon number in the cavity. [7]

$$\omega_r = \sqrt{A\left[\frac{1 + \beta S_0}{\tau_p} - A N_0 \beta (1 - \beta)\right]} \tag{18}$$

Where "$N_0$" is the steady-state population inversion number of carriers. [7]

$$\omega_p = \frac{1}{\tau_p} + A\left(1 - \beta N_0 + \beta S_0\right) \tag{19}$$

As a result, by putting these time constants in (15) effective modulation bandwidth can be achieved and thus the proposed device can perform up to 178GHz.

## 4. Results and output characteristics

One of the most important characteristics of a laser is the output power profile vs normalized pump rate (P/P$_{th}$) which is shown in Fig.12. The behavior of this profile demonstrates the proper laser operation of the introduced structure.

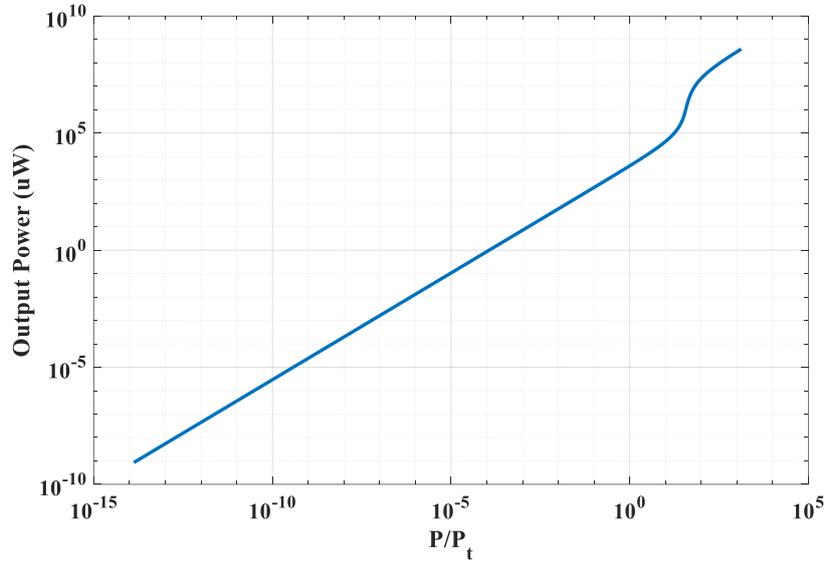

Fig.12 Output power (μW) vs normalized pump rate

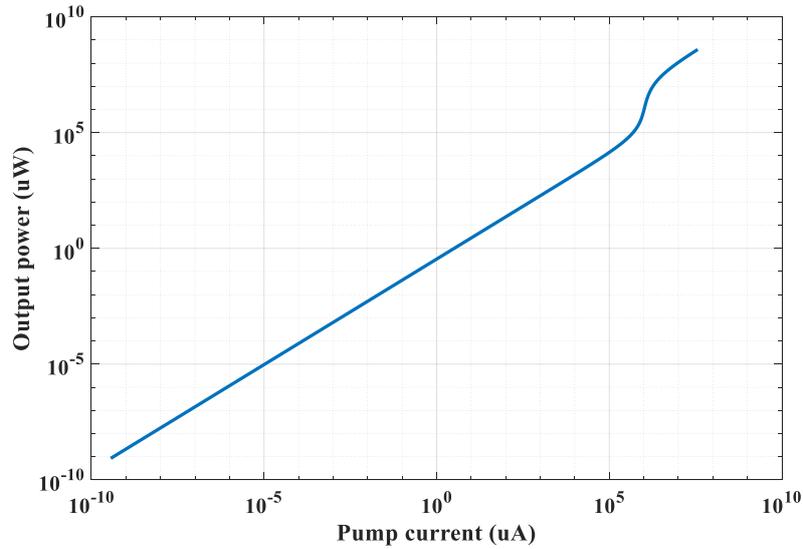

Fig.13 Output power (µW) vs injected current (µA)

In Fig.13 a better input-output characteristic that relates the output SPP power (µW) to the input injection current (µA) is shown. Relatively large output power levels while maintaining the input pump current in microampere levels and in the room temperature, results in a practically appropriate device for chip-level integration processes.

As mentioned before, the threshold current is about 29mA. Due to the tiny area of this device for the threshold current the related current density is about 40000 kA/cm$^2$ which is far greater than the traditional III-V diode lasers. [10] However, this enormous current density can be thermally managed by appropriate cooling solutions in order to achieve mW range output power. It should be mentioned that this device even in 10µA pump current which results in 13.7kA/cm$^2$ provides about 2.8µW output power which makes it useful even in very low pump currents without thermal breakdown. Furthermore, in a plasmonic nanolaser due to the strong coupling of the spontaneous emission into the lasing mode even before the threshold current lasing can be achieved.

In addition, laser gain per unit length of the cavity can also be witnessed in Fig.14.

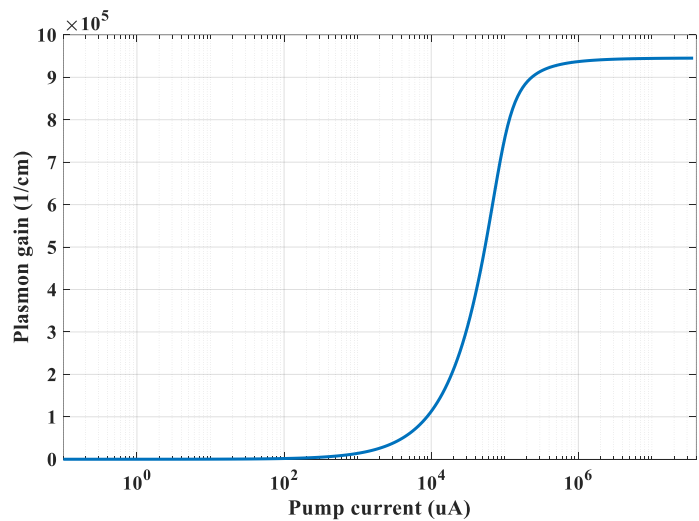

Fig.14 Laser gain (cm$^{-1}$) vs injected current (µA)

Finally, in Table.2 the key parameters of the proposed nanolaser are concluded. Although analysis done in this paper were based on theoretical models, this cannot guarantee if implemented it should work up to the derived performance. Nevertheless, a notable performance can be predicted.

**Table.2 Key parameters of the nanolaser**

| Parameter | This work | Liu, et al [7] |
|---|---|---|
| Output wavelength (nm) | 1550 | 850 |
| Area ($\mu m^2$) | 0.073 | 0.06 |
| Threshold current (mA) | 29 | 0.0019 |
| Output power in mW @ threshold | 4.16 | 0.08 |
| Output power in $\mu W$ @ $10 \mu A$ | 2.8 | 0.25 |
| Gain in $cm^{-1}$ @ threshold | 26680 | - |
| Bandwidth in GHz @ threshold | 178 | > 80 |
| Purcell factor (Lasing mode) | 291 | 15 |
| Coupling factor ($\beta$) | 0.0685 | 0.55 |

## 5. Conclusion

In this paper a Ge/SiGe multi-quantum well metal strip plasmon source is introduced, theoretically analyzed, and numerically simulated. We have confirmed our calculations by means of various physical models and simulation tools. The key advantages of the proposed structure are its tiny footprint ($0.073 \mu m^2$), Silicon friendly process, room temperature operation, electrically pumping and high-efficiency coupling with plasmonic waveguides, which makes it a proper choice for the plasmon source in the development of plasmonic integrated circuits. The new structure generates $2.8 \mu W$ output power with $10 \mu A$ injection current in 1550nm-free space wavelength, has a wide modulation bandwidth of 178GHz, large Purcell factor about 291 which is resulted by considerable mode confinement.